\documentclass[runningheads]{llncs}
\usepackage[T1]{fontenc}
\usepackage{graphicx,verbatim}
\usepackage{amsmath}
\usepackage{amssymb}
\usepackage{multirow}
\usepackage{booktabs}
\usepackage{marvosym}
\usepackage{url}
\usepackage{caption}
\usepackage{enumitem}
\usepackage{hyperref}

\usepackage{xcolor}

\definecolor{addgreen}{rgb}{0,0.5,0}

\begin{document}
\title{Primitive Representation Learning for Unsupervised Dynamic Contrast Enhanced \\ MRI Reconstruction}
\titlerunning{Primitive Representations for DCE MRI}
%
\author{
Veronika Spieker\inst{1,2,3} \and
Wenqi Huang\inst{2} \and
Cemre Ariyurek\inst{3} \and
Liam Timms \inst{3} \and
Daniel Rueckert \inst{2,4,5} \and
Onur Afacan \inst{3} \and
Julia A. Schnabel \inst{1,2,4,6} \and
Sila Kurugol \inst{3}
}
\authorrunning{Spieker et al.}
%
\institute{
Institute of Machine Learning in Biomedical Imaging, Helmholtz Munich, Germany
\and
 School of Computation and Information Technology, Technical University of
Munich, Germany
\and 
Department of Radiology, Boston Children’s Hospital and Harvard Medical School, USA
\and
Munich Center for Machine Learning, Technical University of Munich,
Germany
\and
Department of Computing, Imperial College London, United Kingdom
\and
School of Biomedical Engineering and Imaging Sciences, King’s College London, United Kingdom
\\
\email{v.spieker@tum.de}}

\maketitle              
\begin{abstract}
Reliable quantitative analysis of dynamic contrast-enhanced MRI requires high-quality spatiotemporal reconstructions at high undersampling rates. Scan-specific reconstructions using Gaussian and Gabor primitives have shown promising results without the need for large training datasets, but have not addressed the additional dimension of dynamic contrast. We propose a multi-dimensional, primitive based framework for dynamic contrast-enhanced MRI reconstruction that disentangles the underlying anatomy, the dynamic contrast enhancement, and residual motion into separate temporal basis functions, thereby enabling a geometrical interpretation of the representation. We show that this architecture achieves performance competitive with conventional reconstruction methods, both in reconstruction quality and in the accuracy of extracted aorta and kidney enhancement curves. The modular tier design extends naturally to additional dynamic factors and higher acceleration rates. Code available at \url{https://github.com/compai-lab/2026-GaborDCE-spieker}.%

\end{abstract}

\section{Introduction}

Magnetic resonance imaging (MRI) offers high diagnostic value, but its long acquisition times remain a fundamental limitation. Therefore, acceleration strategies are widely employed, although they often involve a trade-off between spatial resolution, temporal resolution and temporal signal fidelity. This trade-off is especially consequential in dynamic contrast-enhanced (DCE) MRI, where preserving the peak and shape of the arterial input function (AIF) is essential for reliable quantitative parameter estimation, e.g., in MR urography for renal function assessment~\cite{GrattanSmith_2022,Gunwhy_2026,Riffel.2016,Sourbron_2013}.
Standard clinical reconstructions such as GRASP and XD-GRASP~\cite{Feng_2016} prioritize diagnostic image quality, often over-smoothing temporal dynamics and degrading downstream parameter estimation.

Recently, unsupervised deep learning methods have shown promising results for spatiotemporal reconstruction \cite{Heckel_2024,Spieker_2023}. Rather than relying on large training datasets, these approaches fit continuous representations on a scan-specific basis. This is particularly advantageous for abdominal MRI, where such datasets are difficult to obtain, as acquiring motion-free, fully sampled reference data is especially challenging.
One widely applied architecture for spatiotemporal reconstruction is based on implicit neural representations (INRs), which use an MLP to encode intensity values at arbitrary spatial locations (pixels in 2D or voxels in 3D) \cite{Catalan,Feng_2025,Huang_2023,Spieker}. 

However, as purely coordinate-based functions optimized without an explicit spatial prior, INRs are prone to sub-resolution overfitting and offer limited interpretability. Primitive-based methods instead represent the image with an explicit set of kernels whose shape prior removes this failure mode. More recently, a novel representation based on Gaussian primitives has been proposed. Inspired by Gaussian splatting, the image space is parametrized through the properties of multiple Gaussian kernels, namely their weights and geometric parameters, which define their spatial extent and contribution. The final image is rendered by summing all primitives at the desired image coordinates \cite{zhang2024gaussianimage}. Initial applications of primitive representations to MR reconstruction have covered MR slice-to-volume reconstruction (SVR) \cite{Dannecker} and dynamic cardiac MR reconstruction \cite{Huang_2026,Terpstra}. Each primitive has an explicit position, scale and frequency, making the representation directly inspectable, and its reduced parameters train faster than an INR MLP~\cite{Huang_2026}.

Primitive-based MR reconstruction performance was further improved by modulating each Gaussian primitive with a frequency, yielding Gabor primitives.
This approach was applied to cardiac cine imaging, leveraging temporal redundancies through low-rank modeling of the temporal dimension. However, such modeling is not sufficient to also capture contrast enhancement dynamics, which can be considered a decoupled dynamic process. Unlike motion, contrast enhancement is neither smooth nor periodic, but instead requires flexible signal modeling that can accommodate sharp transitions, such as those occurring upon bolus arrival.

We propose to adapt the Gabor primitive idea to DCE-MRI, leveraging the inherent temporal redundancy by modeling the dynamic behavior of the primitives. To this end, we extend the motion-aware reconstruction framework with an additional DCE signal tier, which models the rapidly arriving high-intensity changes on top of the base signal representing the general anatomy. To account for these distinct dynamics, each tier is modeled with its own temporal basis, and the temporal regularization is adapted so that the sharp contrast wash-in is preserved rather than smoothed away. Overall, our contributions are three-fold:
\begin{enumerate}[noitemsep, topsep=0pt, leftmargin=1.5em]
    \item we propose the first primitive-based reconstruction method for DCE-MRI;
    \item we present a novel multi-dimensional design and training strategy for a geometrically interpretable, unsupervised primitive-based representation;
    \item we evaluate on in-vivo pediatric abdominal data, covering both reconstruction quality and downstream quantitative parameters as well as geometrical analysis of the learned representation. 
\end{enumerate}

\section{Methods}

\subsection{Primitive Modeling for MR Reconstruction}

As introduced in \cite{Huang_2026}, we model the dynamic MR image across $T$ frames as a sum of $N$ Gabor primitives whose weights and geometry may vary across frames:
\begin{equation}
x_t(\vec{r}) = \sum_{n=1}^{N} w_{n,t} \, P_n\!\bigl(\vec{r};\; \vec{\mu}_{n,t},\, \vec{s}_{n,t}, \theta_{n,t}, \vec{\xi}_{n,t}\bigr), \quad t = 1, \dots, T,
\label{eq:image_model}
\end{equation}
where $w_{n,t} \in \mathbb{C}$ is the complex weight and $P_n\!\bigl(\vec{r};\, \vec{\mu},\, \vec{s},\, \theta,\, \vec{\xi}\bigr)$ are anisotropic Gabor primitives with its corresponding geometrical parameters $\vec{\mu}_{n,t}$, scale $\vec{s}_{n,t}$, rotation $\theta_{n,t}$ and Gabor frequency $\vec{\xi}_{n,t}$ (see \cite{Huang_2026} for detailed definition).

\subsection{Spatiotemporal Forward Model}
Originally, temporal variation is modeled through two low-rank temporal bases: a \emph{contrast basis} modeling signal-intensity variations, and a \emph{geometry basis} capturing cardiac motion \cite{Huang_2026}. In the following, we expand the model to account for dynamic contrast signal enhancement.
In DCE-MRI, the injected contrast produces a sharp enhancement at time of injection, followed by a slow decay. This temporal behavior is different from the general (non-contrast) anatomy changes, e.g., due to motion, which motivates modeling it as a separate tier. At the same time, the signal enhancement is spatially constrained by the underlying anatomical regions. Assuming that a well-fitted primitive model captures this structure geometrically, contrast enhancement would be expected to manifest simply as an increase in the intensity weight of the corresponding primitives. 

\begin{figure}[t]
\includegraphics[width=\textwidth]{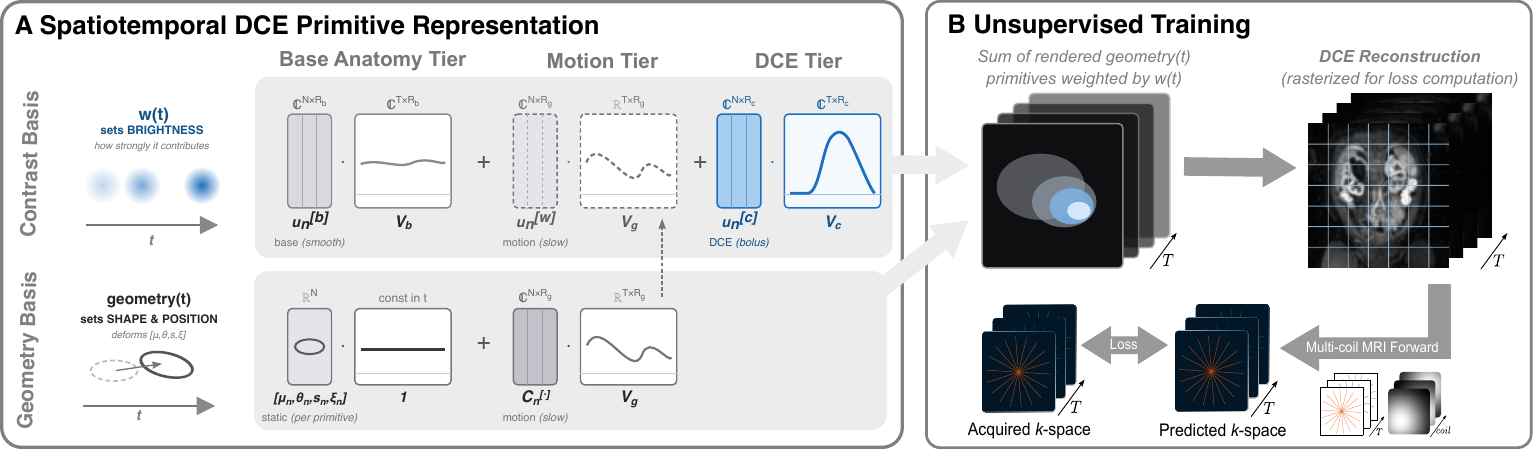}
\caption{Method Overview. (A) Spatiotemporal DCE Primitive Representation consists of a contrast basis and geometry basis, modeling the primitives intensity and the primitives shape/position, respectively. (B) The weighted sum of all rendered primitives results in the DCE reconstruction time series, which is transformed into k-space to compute the unsupervised loss to the acquired data.} \label{fig:overview}
\end{figure}

\subsubsection{Contrast Basis.}
Therefore, we model the complex signal intensity weight $w_{n,t}$ as a composition of three tiers (see Fig.~\ref{fig:overview}A): A \textit{base anatomy} signal representing the pre-contrast anatomy, a  \textit{motion-coupled} signal accounting for motion-induced intensity changes, and a dedicated \textit{dynamic contrast} signal capturing the enhancement and its subsequent decay. As each tier exhibits a distinct temporal behavior, all are modeled with their own temporal basis, resulting in the following summation for the overall intensity weight $w_{n,t}$:
\begin{equation}
w_{n,t} = \underbrace{\vec{u}_n^{[b]} \cdot \vec{v}_{b,t}}_{\text{base tier}}
        \;+\; \underbrace{\vec{u}_n^{[g]} \cdot \vec{v}_{g,t}}_{\text{motion tier}}
        \;+\;  \textcolor{blue}{\underbrace{\vec{u}_n^{[c]} \cdot \vec{v}_{c,t}}_{\text{\textcolor{blue}{DCE tier}}}}
        \;\;,
\label{eq:weights}
\end{equation}
where $\vec{v}_{b,t}\in\mathbb{C}^{R_b}$, $\vec{v}_{g,t}\in\mathbb{C}^{R_g}$ and
$\vec{v}_{c,t}\in\mathbb{R}^{R_c}$ are the $t$-th rows of the shared base, motion and contrast bases $\mathbf{V}_b\in\mathbb{C}^{T\times R_b}$, $\mathbf{V}_g\in\mathbb{C}^{T\times R_g}$ and
$\mathbf{V}_c\in\mathbb{C}^{T\times R_c}$, 
and $\vec{u}_n^{[b]}$, $\vec{u}_n^{[g]}$
and $\vec{u}_n^{[c]}$ are per-primitive
coefficient vectors.

\subsubsection{Geometry Basis.}
The parameters defining the geometry of the primitives ($\vec{\mu}_{n,t}$, scale $\vec{s}_{n,t}$, rotation $\theta_{n,t}$ and Gabor frequency $\vec{\xi}_{n,t}$) are each composed of a static base component and a temporally varying component modeling motion:
\begin{equation}
\begin{aligned}
\vec{\mu}_{n,t} &= \vec{\mu}_n + \mathbf{C}_n^{[\mu]}\vec{v}_{g,t}, &
\theta_{n,t} &= \theta_n + \vec{C}_n^{[\theta]} \cdot \vec{v}_{g,t},\\[2pt]
\log\vec{s}_{n,t} &= \log\vec{s}_n + \mathbf{C}_n^{[s]}\vec{v}_{g,t}, &
\vec{\xi}_{n,t} &= \vec{\xi}_n + \mathbf{C}_n^{[\xi]}\vec{v}_{g,t},
\end{aligned}
\label{eq:geometry}
\end{equation}
where $\log$ denotes the element-wise logarithm and $\mathbf{C}_n^{[\cdot]}$ are per-primitive coefficient matrices for the motion
track. Reusing the geometric temporal base $\vec{v}_{g,t}$ ties
the intensity enhancement and the geometric deformation of an enhancing primitive
to a single time course.

\subsection{Multi-dimensional training strategy}

The primitives are fit by minimizing a density-compensated k-space objective,
\begin{equation}
\mathcal{L}=\sum_{c,t}\big\lVert\sqrt{d}\odot\big(\mathcal{A}_t(S_c\odot x_t)-y_{c,t}\big)\big\rVert_2^2
\;+\;\sum_{i\in\{b,g,c\}}\lambda_i\,\mathcal{R}_i(\mathbf{V}_i)
\;+\;\lambda_x\sum_{t}\big\lVert x_{t+1}-x_t\big\rVert_1,
\label{eq:loss}
\end{equation}
where $y_{c,t}$ are the measured radial samples of coil $c$ at frame $t$, $d$ the density-compensation weights, $\mathcal{A}_t$ the per-frame encoding operator, $S_c$ the coil sensitivity, and $x_t$ the reconstructed frame. The regularizer $\mathcal{R}_i$ is a temporal regularization penalty applied to each tier's temporal basis $\mathbf{V}_i$, with $i\in\{b,g,c\}$ indexing the base, geometry, and contrast tiers (detailed below). The final term is an image-domain temporal total variation, an $\ell_1$ penalty on the frame-to-frame difference of the rendered images $x_t$, weighted by $\lambda_x$.

\noindent\textbf{Temporal regularization.}
We penalize each basis's frame-to-frame differences with a Huber kernel $H_\delta$ to suppress temporal noise while preserving the sharp bolus arrival: differences below $\delta$ are treated quadratically (damped more strongly than $\ell_1$), while the large bolus jump stays in the linear ($\ell_1$) regime and is left un-smoothed. For each tier $i\in\{b,g,c\}$ we apply $\mathcal{R}_i(\mathbf{V}_i)=\bar V_i^{-1}\,\langle H_\delta(\Delta_t\mathbf{V}_i)\rangle$, where $\bar V_i=\langle|\mathbf{V}_i|\rangle$ normalizes by each tier's scale so one $\delta$ yields equal \emph{relative} smoothing across tiers of differing magnitude. All three bases share this penalty; only the weights differ, with the base and geometry bases weighted more strongly than the contrast basis ($\lambda_c, \lambda_g \ll\lambda_b$) to keep the enhancement sharp.

\noindent\textbf{Staged optimization.}
Optimizing the three tiers of Eq.~\eqref{eq:weights} jointly from scratch is unstable because the high-capacity base tier tends to absorb the contrast enhancement that should be captured by the dedicated contrast tier, collapsing the intended separation. We therefore adopt a staged schedule in the following order: base/pre-contrast $\rightarrow$ DCE contrast $\rightarrow$ motion-geometry. 
First, the static base tier $\mathbf{V}_s$ is fit on the pre-contrast frames with the other tiers frozen, so it commits to the pre-enhancement anatomy. The contrast tier is then unfrozen over all $T$ frames, forcing the bolus wash-in into $\mathbf{V}_c$, and finally the motion-coupled geometry $\mathbf{V}_g$ is released to absorb residual deformation. This sequential unfreezing keeps the anatomy, contrast, and motion tiers from competing for the same signal while the separation is most fragile.

\section{Experimental Setup}

\begin{figure}[!t]
\includegraphics[width=\textwidth]{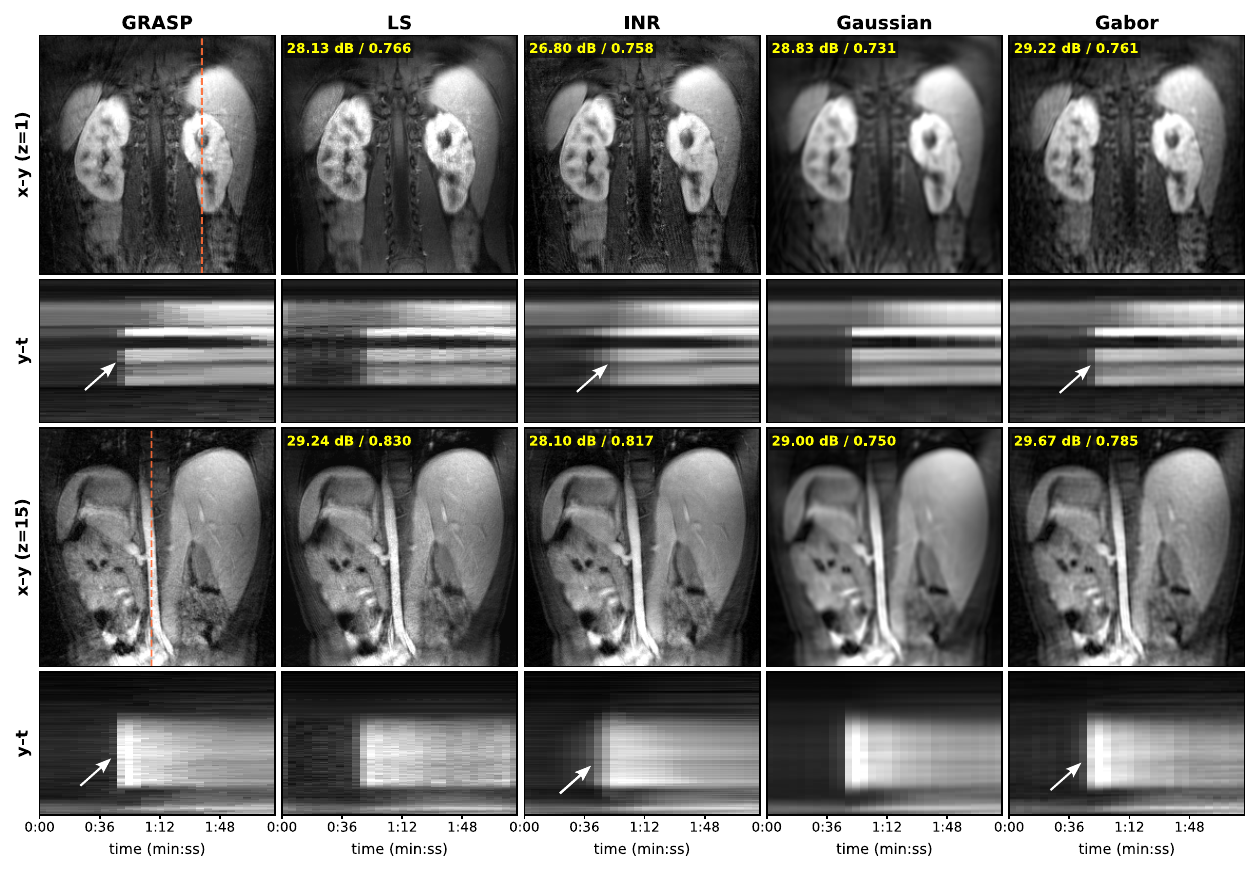}
\caption{Qualitative spatial ($xy$) and spatiotemporal ($xt$) reconstruction results for an exemplary subject (kidneys/aorta on the top/bottom, respectively).  While INR results in competitive but slightly noisy reconstructions compared to the conventional references, it blurs spatiotemporally (arrows). Gabor sharpens reconstructions compared to Gaussian and maintains temporal sharpness.} \label{fig:Reconstruction}
\end{figure}


\subsection{Data}

We evaluate on five in-house pediatric MRI acquisitions using a golden angle stack-of-stars 3D FLASH prototype sequence with a multichannel body-matrix coil (3T Siemens VIDA/PRISMA) to acquire 32 coronal slices with a voxel size of $1.25\times1.25\times3\,\mathrm{mm}^3$, TR/TE/FA $=3.56/1.39\,\mathrm{ms}/12^\circ$. The acquired spokes were retrospectively binned to 34 spokes per frame, resulting in a temporal resolution of $\sim$4.2-4.8 seconds per volume. For an image size of $448{\times}448$ pixels, this corresponds to an undersampling factor of $R\approx21$ relative to the Nyquist criterion. Reconstructions were performed on all slices for the complete field-of-view and then cropped to a $224{\times}224$ region of interest. To reduce computational overhead, coil compression was applied to retain 80\% of the coil information, and coil sensitivities were estimated via ESPIRiT~\cite{uecker2014espirit}.
Aorta/kidney segmentations were obtained with nnU-Net~\cite{Isensee} and verified manually. 

\subsection{Evaluation Methods}
Methods are qualitatively compared on representative aorta and kidney slices, both spatially and spatiotemporally. Since no ground truth is available, we compute PSNR and SSIM against GRASP as an indicator of reconstruction similarity. To assess the clinical potential for urography, including tracer kinetic modeling~\cite{Sourbron_2013}, we further analyze the temporal behavior of the aorta and kidney in the spatiotemporal reconstruction. To this end, we reconstruct the whole volume and extract the aorta and kidney signal intensities using segmentation masks, retaining the top 20\% of signal intensities within each mask. Additionally, we quantify the temporal sharpness relevant for AIF estimation by computing the peak-to-baseline ratio (P/B), the wash-in slope and the temporal variation after contrast arrival ($TV_{post}$).
Comparison methods include the current clinical standard \textbf{GRASP} \cite{Feng_2016} with low temporal regularization used for accurate kidney function quantification, \textbf{L+S} \cite{Otazo_2015} as a conventional decomposition-based method, and \textbf{Hash-INR} as an alternative scan-specific learning-based method, based on an implicit neural representation with multi-resolution hash-grid encoding~\cite{muller2022instant}. Also, a \textbf{Gaussian} primitive representation ($\xi_n=0$) is computed.

\subsection{Implementation Details}
All methods were implemented in PyTorch on an NVIDIA RTX A5000 GPU. Custom CUDA kernels were used for primitive rasterization. Our reconstruction is initialized with 15000 Gabor primitives on a regular grid and adaptively densified up to 20000 during optimization (19000 and 25000 for Gaussian primitives to match parameter count, respectively). Adaptive density control prunes low-contribution primitives and splits/clones high-gradient ones every $300$ iterations, active between $25\%$ and $50\%$ of training. The base, contrast and geometry tiers use temporal ranks $R_b{=}3$, $R_c{=}4$ and $R_g{=}1$. All parameters are optimized with Adam for 5000 iterations under a cosine learning-rate schedule, with per-group rates.
Optimization begins with 500 pre-contrast iterations on the static tier alone, followed by 500 iterations with the DCE tier unfrozen, before the full model is optimized from iteration 1000 onward. Huber total-variation ($\delta{=}0.05$) is weighted by $\lambda_{b}, \lambda_g = 5{\times}10^{-4}$ and $\lambda_{c} = 1{\times}10^{-4}$, and $\lambda_{x}$ is set to $1{\times}10^{-3}$.
GRASP reference reconstructions were obtained using a public implementation \cite{Ariyurek_2023_grasp}, with a relative temporal lambda of 0.05. 
L+S regularization parameters were set to $\lambda_L=0.01 / \lambda_S=0.005$.
Hash-INR (16 hash levels and $2^{19}$ map size for encoding, 2-hidden-layer MLP with 128 nodes) was trained for 5k iterations.

\section{Results}
\begin{figure}[t]

\centering
\captionof{table}{Aorta TIC metrics. Mean $\pm$ std across 5 subjects (ROI-averaged).} \label{table:tic_measures}
\begin{tabular}{lccccc}
\toprule
Metric & GRASP & LS & INR & Gaussian & Gabor \\
\midrule
P/B $\uparrow$ & $3.50 \pm 0.84$ & $2.89 \pm 0.62$ & $3.01 \pm 0.58$ & $3.41 \pm 0.83$ & $3.37 \pm 0.78$ \\
wash-in $\uparrow$ & $0.35 \pm 0.15$ & $0.17 \pm 0.06$ & $0.12 \pm 0.03$ & $0.30 \pm 0.13$ & $0.29 \pm 0.13$ \\
TV$_{\text{post}}$ $\uparrow$ & $0.15 \pm 0.05$ & $0.12 \pm 0.05$ & $0.09 \pm 0.03$ & $0.14 \pm 0.04$ & $0.14 \pm 0.04$ \\
\bottomrule
\end{tabular}

\vspace{1em}

\includegraphics[width=\textwidth]{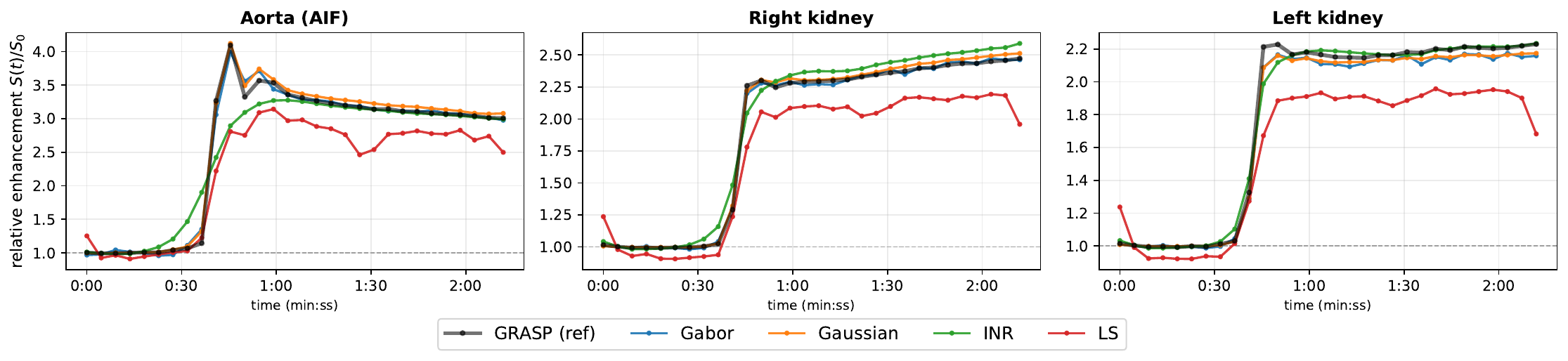}
\captionof{figure}{Aorta and kidney time intensity curve comparison. A sharp aortic peak is expected as soon as the contrast is injected ($\sim$35\,seconds). INR over-smooths the signal, while Gaussian and Gabor maintain a sharper peak than the reference GRASP. Kidney curves confirm similar dynamic behavior.} \label{fig:aorta_tic}

\end{figure}

\noindent\textbf{Reconstruction quality.}
Figure~\ref{fig:Reconstruction} shows spatial ($xy$) and spatiotemporal ($xt$) reconstructions for two slices. Hash-INR is competitive spatially but slightly noisier than the GRASP reference and blurs along the temporal dimension. Among the primitive models, Gabor recovers sharper spatial detail than the Gaussian variant and preserves the temporal sharpness that the INR loses. 
Quantitative image similarity confirms the sharper reconstruction, yet these should be regarded as indicative only, since GRASP does not constitute a true ground truth.

\noindent\textbf{Time-intensity curves.}
Figure~\ref{fig:aorta_tic} compares aorta and kidney time-intensity curves; a sharp aortic peak is expected shortly after injection (${\sim}35$\,s). Hash-INR and L+S oversmoothen this peak, whereas the Gaussian and Gabor reconstructions retain a peak similar to the GRASP reference, and the kidney curves show the same behaviour. Quantitative values (Table~\ref{table:tic_measures}) show similar peak behavior as well as post contrast temporal smoothness of GRASP, Gabor and Gaussian.

\begin{figure}[t]
\includegraphics[width=\textwidth]{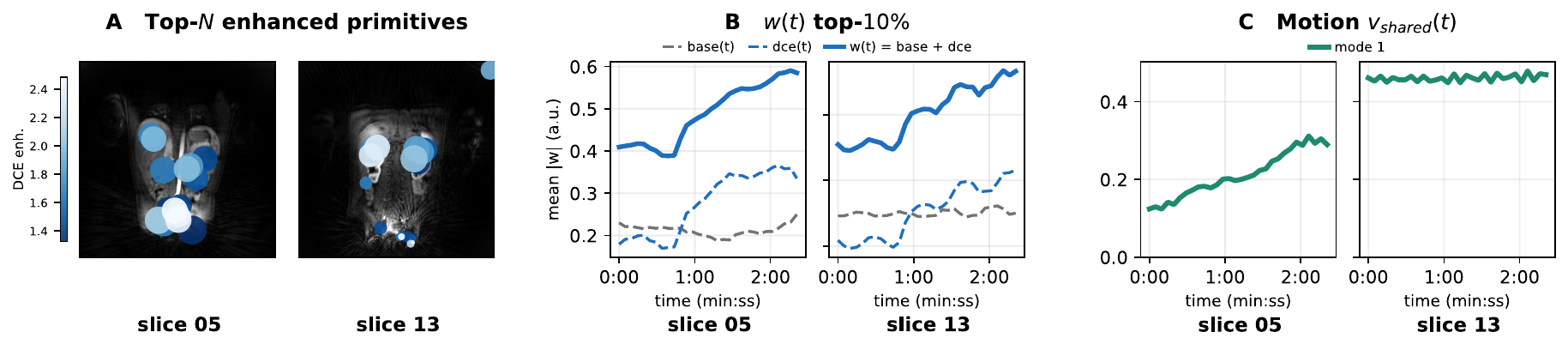}
\caption{
Gabor representation analysis. (A) Location and scale of the top-20 enhanced primitives overlaid on the reconstruction. (B) Base- and DCE-tier contrast intensity over time for the top-10\% enhancing primitives; the base stays flat while the DCE tier rises at injection. (C) Learned motion-tier temporal basis (general drift, slice 05; respiratory motion, slice 12).
} \label{fig:dce_analysis}
\end{figure}

\noindent\textbf{Representation analysis.}
Figure~\ref{fig:dce_analysis} inspects the learned representation for a subject. Panel~(A) overlays the most enhancing primitives on the reconstruction, indicating that their locations coincide with the expected enhancing anatomy. Panel~(B) shows the contrast intensity contributed by the base and DCE tiers over time for these primitives. The base tier stays flat while the DCE tier rises at injection, showing that the model separates the contrast dynamics from the static anatomy. Panel~(C) visualizes the learned motion basis, which exhibits a periodic pattern for slice~13 that may be attributed to respiratory motion.

\section{Discussion and Conclusion}

We presented the first primitive-based framework for DCE-MRI reconstruction. By disentangling the representation into a base, a dynamic-contrast, and a geometry-coupled tier, the model reconstructs highly undersampled acquisitions with quality competitive with the clinical GRASP reference, while preserving the sharp aortic and renal enhancement that the comparison methods over-smooth. The recovered contrast primitives localize to the expected anatomical regions and separate the enhancement from the static anatomy, 
yielding a direct geometric interpretation. The learned motion basis further indicates that a primitive representation can jointly capture contrast and motion in a disentangled manner.

Yet, the present evaluation is limited by the absence of a ground-truth reference: the quantitative comparison is made against GRASP, which itself temporally smooths the bolus, and is restricted to a small in-vivo cohort. Rendering every frame also couples the reconstruction cost to the temporal resolution, trading training time for finer dynamics and requires further hyperparameter tuning. 

Overall, we have shown the potential of representing undersampled DCE-MRI with an explicit geometric meaning, a representation that inherently supports higher acceleration through explicit regularization. Extending the framework to further multi-dimensional MRI settings and driving downstream tasks such as tracer-kinetic parameter estimation motivate further exploration.

%
%
%
\bibliographystyle{splncs04}
\bibliography{references}
%
\end{document}